\documentclass[runningheads]{llncs}
\usepackage{graphicx}
\usepackage{color}
\usepackage[hyphens]{url}

\begin{document}
\title{Can an Algorithm be My Healthcare Proxy?}
%
%
\author{Duncan C McElfresh\inst{1} \and
Samuel Dooley\inst{1} \and
Yuan Cui\inst{2} \and
Kendra Griesman\inst{3} \and
Weiqin Wang\inst{4} \and
Tyler Will\inst{5} \and
Neil Sehgal\inst{1} \and
John P Dickerson\inst{1}
}

\authorrunning{D.C. McElfresh et al.}
%
\institute{University of Maryland, College Park\\
\email{dmcelfre@umd.edu} \and
Oberlin College \and
Haverford College \and
Pennsylvania State University\and
Michigan State University
}
\maketitle              
\begin{abstract}
Planning for death is not a process in which everyone participates. Yet a lack of planning can have vast impacts on a patient's well-being, the well-being of her family, and the medical community as a whole. Advance Care Planning (ACP) has been a field in the United States for a half-century.
Many modern techniques prompting patients to think about end of life (EOL) involve short surveys or questionnaires.
Different surveys are targeted to different populations (based off of likely disease progression or cultural factors, for instance), are designed with different intentions, and are administered in different ways.
There has been recent work using technology to increase the number of people using advance care planning tools. 
However, modern techniques from machine learning and artificial intelligence could be employed to make additional changes to the current ACP process.
In this paper we will discuss some possible ways in which these tools could be applied. 
We will discuss possible implications of these applications through vignettes of patient scenarios.
We hope that this paper will encourage thought about appropriate applications of artificial intelligence in ACP as well as implementation of AI in order to ensure intentions are honored. 

\end{abstract}

\section{Introduction}

Modern medicine is exceptional at prolonging life. New devices, drugs, and medical procedures can stave off death in cases that--until recently--were immediately fatal. Legal and ethical norms dictate that each patient may decide \emph{for themselves} whether or not to receive these life-sustaining treatments (LST); however patients in need of EOL care are often physically or cognitively unable to make these decisions. In anticipation of this, some people use Advance Care Planning (ACP) in order to better understand their care options and ensure they receive only the care they want. Medical decision-making is challenging in general, and each patient's unique values and health conditions complicate things further. ACP is often an iterative process with many conversations--involving the patient's family, medical professionals, and sometimes ethicists, lawyers, and religious representatives. In the best case, ACP culminates in a written description of the patient's wishes, such as an Advance Directive (AD) or Physician Orders for Life Sustaining Treatment (POLST); these documents are intended to represent the patient's wishes to their family and medical professionals, when the patient is unable to do so.  

However, most people do not participate in ACP; in the US, completion rates are especially low among less-educated, poorer, and non-white people~\cite{rao2014completion}. Without ACP, EOL care decisions are often left to a proxy decision maker--usually a family member or friend. It is a well-known, tragic fact that proxies often make decisions that contradict their patient's wishes. Usually this means providing LST until it becomes medically or financially impossible. Insufficient care planning has ethical, legal, and financial implications; LST is often inconsistent with patients' authentic wishes, expensive, and emotionally and morally burdensome for the proxy decision maker. Some of these problems persist even \emph{with} ACP since existing forms do not address all EOL care scenarios, and may not lead to preferred outcomes.

With Artificial Intelligence (AI)--and in particular  Machine Learning (ML)--being adopted in various medical and health fields, we pose the question: what role \emph{can} and \emph{should} AI play in ACP? We begin in Section~\ref{sec:history} with a brief history of ACP, including modern automated and web-based applications. Then in Section~\ref{sec:forward} we look forward to new AI methods that might soon play a role in ACP. Indeed, the medical community already employs AI for a variety of tasks; it is easy to imagine such an application for ACP. In Section~\ref{sec:apps} we discuss two hypothetical applications in this space and discuss their implications. We conclude in Section~\ref{sec:conclusion} with a high-level discussion and recommendations for future work.

\section{A Brief History of Advance Care Planning}\label{sec:history}

Formal ACP was first proposed in 1967~\cite{glick1991right}, based on the legal premise that patients cannot be subjected to medical treatment without consent. This gave way to the first formal ACP documentation, a \emph{living will} (LW), in which patients document exactly when to provide--and when to withhold--LST. LWs were rapidly accepted across the US; by 1986, 41 states had enacted LW statutes~\cite{glick1991right}. The limitations of LWs became rapidly apparent: their directives were often too vague, or too particular, to influence care decisions.

In response, policymakers created a mechanism allowing patients to appoint a \emph{medical power of attorney} (MPOA)\footnote{Nomenclature varies by state and may be phrased as Durable Power of Attorney for Health Care, Health Care Proxy, Surrogate Decision Maker}, often a family member or close friend, to make care decisions on their behalf when they could not do so themselves~\cite{sabatino1991death}. Around the same time, states began enacting protocols to recognize Do Not Resuscitate (DNR) and Do Not Intubate (DNI) orders.

The latest innovation in ACP is the POLST, a process for translating a patient's goals and wishes into concrete medical guidance~\cite{bomba2012polst}. POLSTs involve in-depth conversations between a patient and her provider, and are meant to go beyond the rigid structure of ADs to capture the patient's unique health conditions and goals. 

Modern ADs include most of the components described above: precise LW statements for EOL care, DNR/DNI statements under varying circumstances, and general wishes or treatment goals. Yet modern methods have still not fulfilled the lofty and important goals of ACP~\cite{clemency2017decisions,hickman2017quality}.

Two primary challenges persist: (1) it is difficult to translate patients' unique preferences and goals into care decisions, and (2) few people actually participate in ACP (30\% of adults in the US~\cite{rao2014completion,yadav2017approximately}). To address these challenges, some have turned to automation and the internet.

\subsection{Computer-Based ACP}

While ACP is traditionally a human-focused endeavor, many computer- and web-based applications have been developed in recent decades. 
These applications range in their focus and function: some simply provide information, while others help patients complete ADs; some are illness-specific, while others are more general~\cite{butler2014decision}. We outline two popular applications here:\\

\noindent\textbf{PREPARE}\footnote{{\url{https://prepareforyourcare.org/}}} helps elderly users effectively identify and communicate their wishes for care. This application uses several information sources such as testimonials, videos, and narratives to mitigate cultural and communication barriers. PREPARE does not create an AD, but provides a printable ``action plan'' recording the user's wishes. According to the app's creators, PREPARE leads to greater rates of documentation and engagement than traditional methods~\cite{sudore2018engaging}. \\

\noindent\textbf{MyDirectives}\footnote{\url{https://mydirectives.com//}} allows users to create and update a digital AD. Users are asked a series of questions about values and goals; many of these questions are supplemented with clarifying information on the relevant medical procedures or health conditions. The app's creators claim their application enables patients to create a more nuanced AD, and is also more accessible to a wider audience than traditional methods~\cite{fine2016early}. 
\vspace{1em}

Existing applications such as PREPARE and MyDirectives have been shown to help engage people in ACP. But modern ACP applications still fall short: they do not capture patients' nuanced values and wishes, and often fail to influence care decisions~\cite{sabatino2014advance}. With these challenges in mind, we now turn to the future of AI and ACP.

\section{Looking Forward: An ACP Decision Aid}\label{sec:forward}

Computers and modern AI techniques have been shaping the medical profession in myriad ways. Early examples include decision-aids for diagnosis and treatment planning~\cite{shortliffe1979knowledge,kahn1996decision}; modern ML methods have been broadly applied to detect cancers~\cite{kourou2015machine}; automated alerts are widely used to detect drug interactions~\cite{jung2013attitude} and sepsis~\cite{sawyer2011implementation}.

Future applications for ACP will likely take the form of \emph{decision aids}, designed to help care providers make better decisions for their patients. Computerized decision aids have been developed for a variety of purposes including diagnosis, prevention, disease management, and prescription management; however their impact on patient outcomes is not always clear~\cite{garg2005effects}. These applications leverage expert knowledge to make predictions (e.g., a diagnosis) or recommendations (e.g., a treatment) to a care provider. In ACP, the decision question is often whether or not to provide LST, and for how long and in what forms. Answering these questions is not a matter of expert knowledge, but rather one of patient goals and wishes. And patient preferences are not simple: they depend on a variety of factors, such as personal values, religious beliefs, goals of care, cultural background, and family preferences ~\cite{patient2019milnes,gramling2019end,lee2013exploring,patient2013winter}. 

An effective ACP decision aid would understand, and accurately represent, a patient's wishes for care. This is a complicated task, but recent advancements in computational methods for preference elicitation and recommender systems could provide innovation. We briefly describe these fields here.

\subsubsection{Preference Elicitation} is the study of people's preferences, usually by learning a \emph{utility function}. This field has its roots in marketing and economics though recent applications include healthcare~\cite{llewellyn2013decision} and public policy~\cite{alvarez2002using}. The AI community has developed a wealth of elicitation methods which can be applied to a huge variety of scenarios (e.g., see~\cite{boutilier2002pomdp,ailon2012active}).

\subsubsection{Recommender Systems} share a similar lineage with preference elicitation. Used mostly in commercial settings, recommender systems use consumer data to suggest products (e.g., Amazon), content (e.g., Netflix), or social connections (e.g., Facebook) to users; for a review, see~\cite{lu2015recommender}.

\vspace{1em}
Preference elicitation is the study of \emph{what people want}; once a person's preferences are known (or inferred), recommender systems identify an action or outcome consistent with their preferences. We anticipate that an ACP decision aid would leverage similar techniques to learn--allowing care providers to take action on--a patient's care preferences. 

Every AI application has risks. Automating healthcare can threaten patient autonomy, privacy, and can worsen existing disparities in healthcare~\cite{wsj,gianfrancesco2018potential}. These risks are especially high in ACP, which focuses on life-and-death decisions. Our goal in this paper is to anticipate these applications, and their potential impact on patient outcomes. Next, to spark discussion, we outline two hypothetical AI-based ACP applications.

\section{Applications}\label{sec:apps}

We present two applications with the same goal -- to improve patient care outcomes through ACP. Further, the data and algorithms they use are very similar (if not identical).  However the role they play in ACP is quite different, as are their implications.

\subsection{Application 1: \textsc{CareDecider}}

First we consider \textsc{CareDecider}, an AI-driven healthcare proxy. \textsc{CareDecider} consists of two different web interfaces: one for patients, and one for care providers. Patients create a profile and enter basic demographic and health information; they might be prompted to answer some follow-up questions resembling those on ADs. After answering these questions, each user is instructed to print and sign an MPOA, designating \textsc{CareDecider} as their proxy (in the US this is a matter of state law). As with (human) healthcare proxy forms, patients provide copies to their family and care providers. Most importantly, the \textsc{CareDecider} proxy form provides a unique \emph{patient login code}, which care providers use to access a patient's \textsc{CareDecider} profile. When a provider logs in, they answer several questions regarding the patient's state, prognosis, and the treatment options being considered. Upon entering this data, \textsc{CareDecider} provides an estimate of whether or not the patient would want to receive aggressive LST. Consider the following example:
 
\subsubsection{Example} A middle-aged woman arrives in a hospital's emergency department after suffering a major head injury in a car accident. She is in a coma and unable to communicate, though she is otherwise healthy. Her physician determines that she has almost no chance of cognitive recovery, and will likely require artificial nutrition and hydration (ANH) for the rest of her life if she survives the acute episode. The hospital has a copy of the patient's \textsc{CareDecider} proxy form on file, so the physician logs in to \textsc{CareDecider} using the patient's unique code. The physician answers several questions about the patient's prognosis (coma with little chance of recovery) and the treatment option being considered (ANH versus allowing natural death). Using this information, \textsc{CareDecider} makes an inference about the patient's preferences: when creating her profile, she stated that she is strongly religious; in the \textsc{CareDecider} database, 97\% of other patients who reported being strongly religious also stated they want to receive LST no matter the circumstances. Using this (among other factors), \textsc{CareDecider} states that, with 94\% confidence, the patient wants to receive ANH and other aggressive measures to keep her alive despite her injuries.

\subsubsection{Back-End} Behind the scenes, \textsc{CareDecider} resembles a standard ML application -- predicting patient care preferences using training data. In this case, training data might consist of care decisions made by other patients in the past (from hospital records, or the application itself). 
Engineers might encode these decisions into numerical vectors representing each patient's state and the treatment option considered (i.e., the \emph{input variables}), and whether they decided to receive the treatment or not (i.e., the \emph{output variables}). Using modern ML methods, \textsc{CareDecider}'s engineers train a model that infers patient care preferences (e.g., they learn a \emph{utility function} for the patient). When a user creates a \textsc{CareDecider} profile, they are prompted with additional questions to provide providing personalized training data.

These additional questions might be selected using techniques from active learning (a ML subfield) or preference elicitation. When a care provider logs in and fills out the requested information, they receive an estimation of the patient's preferences.

While apocryphal, \textsc{CareDecider} represents a simple and technically feasible use of AI in healthcare. Deploying this application would only require sufficient training data, and a web interface to initialize and query patient's personalized ML model. From a ML perspective, \textsc{CareDecider} allows care providers to directly query the (approximation of) a patient's utility function. Indeed this is the general structure of hypothetical applications proposed for a similar purpose--predicting patient mortality to facilitate EOL care discussions~\cite{avati2018improving,parikh2019machine}.

In the best case, \textsc{CareDecider} might represent patient goals and wishes more accurately than their AD or (human) proxy. In the worst case, \textsc{CareDecider} might misrepresent a patient's wishes, or violate patient autonomy (e.g., making decisions that the patient never considered). Perhaps the greatest risk is that \textsc{CareDecider} will encode human biases into its predictions, and will disproportionally impact already-disadvantaged groups (e.g. people without health insurance or access to a doctor)~\cite{crawford2016there}. We leave further discussion to future work. While some of these risks are inherent to any AI-based system, others are a matter of design. Next we consider an ACP application that -- using very similar methods to \textsc{CareDecider} -- avoids (or perhaps conceals) some of the apparent risk.

\subsection{Application 2: \textsc{PrefList}} 

Next we consider \textsc{PrefList}, which closely resembles existing web-based applications such as MyDirectives. However unlike these applications, \textsc{PrefList} uses an AI back-end to create a unique list of questions for each patient to answer. The patient-facing interface of \textsc{PrefList} is identical to that of \textsc{CareDecider}: patients enter demographic and health information, followed by some additional questions, similar to those on modern ADs. However the provider-facing interface is far simpler: it displays only the information provided by the patient, including their answers to any AD-style questions. In this way, \textsc{PrefList} fits into the existing field of AD tools, while also leveraging modern AI methods. Consider the following example:

\paragraph{Example} A 73 year-old man with lung cancer in remission creates a \textsc{PrefList} account. Using answers to his initial health questions, \textsc{PrefList} determines that his cancer may return, and this may change his care preferences. In addition to the standard AD questions, \textsc{PrefList} asks the man two additional questions: how would his preferences change if he had a terminal illness and only 1 year to live? He states that his preferences would not change, and \textsc{PrefList} asks if his preferences would change with only 1 month to live.

\paragraph{Back-End} Behind the scenes, \textsc{PrefList} may be identical to \textsc{CareDecider}. However unlike \textsc{CareDecider}, the patient's personalized ML model is only used to select questions--and is not directly available to care providers. Yet this ML model still plays an important role in \textsc{PrefList}, by guiding the interaction between patient and questioner. Methods from active learning, preference elicitation, and recommender systems can still be applied in this setting, with a less-direct impact on patient outcomes.

The benefits of \textsc{PrefList} seem clear: patients can create personalized ACP (similar to POLST), without the assistance of a medical professional. Furthermore, the AI back-end might identify which questions are most important for each patient to answer, given their unique goals and wishes. \textsc{PrefList} doesn't appear to raise any risk to patient autonomy, at least not compared to \textsc{CareDecider}. However many risks inherent to AI might still be present, and less apparent, in \textsc{PrefList}.

In particular, issues of bias would be less apparent, yet still present in an elicitation-focused application such as \textsc{PrefList}. For example, if training data includes only decisions and health conditions of (say) older white men, the utility of this application will be questionable for other groups. Using a biased model, \textsc{PrefList} might guide patients to answer uninformative questions. 

Yet another concern is raised by any interaction of AI systems with humans.  It is not immediately clear how a doctor would perceive or react to a description from a computer about a human's desires. For instance, does the doctor trust the computer more over time and eventually stop questioning its authority? There is some initial evidence that \emph{stated} accuracy--not just observed accuracy--impacts a human’s trust in an AI system~\cite{Yin19:Understanding}, which might also lead to questions of manipulation. On the patient's side, she might form a relationship with a system like \textsc{PrefList} which could develop complacency around her EOL decisions.

\section{Discussion}\label{sec:conclusion}

ACP, by nature, is an interdisciplinary social problem. 
As technology evolves, it becomes natural to ask about the appropriate use of that technology in existing areas. 
In the area of ACP, these questions are just beginning to be asked with respect to AI.
Even the introduction of computers in ACP is a recent development. 
But, to immediately incorporate techniques from artificial intelligence without regard for their implications, is unwise. 
In this paper, we have identified some applications that could possibly be developed; we encourage the the ACP community, medical ethicists, and other AI researchers to discuss their appropriateness.
Future work must be done in ethical, quantitative, and qualitative aspects of this problem. 

Before any technological intervention in peoples' lives, particularly in the medical field, there should be considered attention given to whether that intervention is appropriate. 
To facilitate this dialog, there should be more formal discussion between AI researchers and medical ethicists where collaboration occurs.
Possible questions to be asked here include \emph{whether} if a computer should/could make \emph{health} decisions on behalf of a human, and if so, how will conflicts between a patient’s computer-assisted AD and a human MPOA be resolved?

Additionally, there are further technological research needs.
For instance, there should be a concrete formalism describing potential medical actions and current patient states.
Further, for any proposed technological intervention into the current ACP process, there must be a rigorous method to evaluate what added benefit the AI system has over current survey techniques. 
If the theory behind the use of AI in ACP holds that the surveys used will be more detailed and exhaustive, then does that hold true when compared to other EOL questionnaires (e.g., \cite{downey2009shared,engelberg2006psychometric})?

This paper does not attempt to solve problems in ACP; it aims to open a door to deeper thought on the intersection of AI and ACP.\\

\noindent\textbf{Acknowledgments.}  Dickerson, Dooley, and McElfresh were supported in part by NSF CAREER Award IIS-1846237 and by a generous gift from Google. Cui, Griesman, Wang, and Will were supported via an REU grant, NSF CCF-1852352,
and were advised by Dickerson at the University of Maryland. Many thanks to Patty Mayer, who provided thoughtful guidance from the perspective of clinical ethics.

%
%
\bibliographystyle{splncs04}
\bibliography{refs}

\end{document}